\documentclass[useAMS,usenatbib]{mn2e} 
\usepackage{aas_macros}
\usepackage{graphics}
\usepackage[pdftex]{graphicx}
\usepackage{epstopdf}
\usepackage{epsfig}  
\usepackage{natbib} 
\usepackage{float}
\usepackage{amsmath}
\usepackage{times}
\usepackage[varg]{txfonts}
\usepackage{verbatim} % need that for multi-line comments
\usepackage{multirow,bigdelim} % to do the large brackets in the mock table
\usepackage{color}

\usepackage[normalem]{ulem}

\title[Update on Cosmic-flows HI data]{Update on HI data collection from GBT, Parkes and Arecibo telescopes for the Cosmic Flows project}
\author[Courtois \& Tully]
{
H\'el\`ene M. Courtois$^{1}$ and R. Brent Tully$^2$\\
$^1$University of Lyon; UCB Lyon 1/CNRS/IN2P3; IPN Lyon, France\\
$^2$Institute for Astronomy (IFA), University of Hawaii, 2680 Woodlawn Drive, HI 96822, USA\\
}

\begin{document}
\date{}

\pagerange{\pageref{firstpage}--\pageref{lastpage}} \pubyear{2014}

\maketitle

\label{firstpage}

\begin{abstract}
Cosmic Flows is an international multi-element project with the goal to map motions of galaxies in the Local Universe. Kinematic information from observations in the radio HI line and photometry at optical or near-infrared bands are acquired to derive the large majority of distances that are obtained through the luminosity-linewidth or Tully-Fisher relation.  This paper gathers additional observational radio data, frequently unpublished, retrieved from the archives of Green Bank, Parkes and Arecibo telescopes. Extracted HI profiles are consistently processed to produce linewidth measurements. Our current "All-Digital HI Catalog" contains a total of 20,343 HI spectra for 17,738 galaxies with 14,802 galaxies with accurate linewidth measurement useful for Tully-Fisher galaxy distances. This addition of 4,117 new measurements represents an augmentation of 34\% compared to our last release. 
%The linewidth measurement pipeline "Wmean50" is now publicly available for implementation by larger surveys, for example to analyze upcoming HI data from SKA pathfinders telescopes.
\end{abstract}


\section{Introduction}
\label{sec:intro} 

In the cosmology community there is an increasing interest in measuring peculiar velocities of galaxies to map
cosmic flows in the local volume \citep{Courtois2013,Tully2014Nature}. Large coherent flows can provide a direct definition of the major structures
formed by the total dark and luminous matter without the assumptions associated with interpretation of redshift surveys which are probing only the bright galaxy distribution.
Currently data compilations for flow studies are numerically dominated by galaxy distances derived using HI radio detections and application of the Tully-Fisher relation \citep{2013AJ....146...86T}.  It is of interest that
the recent 6dFGSv survey completed at the Schmidt telescope in Australia
\citep{2014arXiv1406.4867C} is now adding a similar number of Fundamental Plane galaxy distance estimates. 

An associated use of peculiar velocities is for measurements of the local bulk flow and its physical interpretation toward understanding the global motion inferred by the CMB dipole \citep{2009MNRAS.392..743W, Hoffman2014}.

A third use for such surveys is to provide better constraints
on parameters of cosmological interest than available from a survey of redshifts alone \citep{2004MNRAS.347..255B, 2005MNRAS.357..527Z}. 
 Comparisons of the galaxy-galaxy, galaxy-velocity and velocity-velocity power spectra \citep{2014arXiv1404.3799J}
  can define cosmological parameters such as the redshift space distortion and the correlation between galaxies and dark matter  
that would be  degenerate when only the information provided by redshift surveys is used.

New peculiar velocity surveys will be quite competitive as cosmological
probes \citep{2013arXiv1312.1022K}. For example, peculiar velocity can improve the growth rate constraints by a factor two (and up to five) compared to density alone
for surveys with galaxy number density of about $10^{-2}~{\rm Mpc}^{3}$.

Future peculiar velocity surveys such as TAIPAN \citep{2013IAUS..289..319C}, and the all-sky HI surveys, WALLABY and WNSHS \citep{2012PASA...29..359K}, can measure the growth rate,
%f$_{\sigma 8} $ 
to 3 per cent at z $\sim$ 0.025. Although the velocity subsample is about an order of
magnitude smaller than the redshift sample from the same surveys, the added information improves constraints by 40 per cent compared to the same survey without velocity measurements.

Such measurements on large scales at z=0 can detect signatures of modified
gravity or non-Gaussianity through a scale-dependent growth rate or galaxy bias. 

The international project Cosmic Flows has  accumulated data to derive accurate distances from both Tully-Fisher and Fundamental Plane techniques. 
Two catalogs of distances have already been produced \citep{2008ApJ...676..184T, 2013AJ....146...86T} .  For Tully-Fisher purposes, our group has been acquiring linewidths derived from HI 21 cm line profiles resulting from observations with the Green Bank Telescope at the US National Radio Astronomy Observatory and with the 13 channel Multibeam Receiver on the Australian 64m Parkes Telescope \citep{2009AJ....138.1938C,2011MNRAS.414.2005C}. Our observations are complemented by archival material from seven telescopes (GBT, NRAO 140-foot and 300-foot, Arecibo, Parkes, Nancay, Effelsberg).  With the addition of the material to be discussed, now $\sim 18,000$ profiles and their interpretation are publicly available at the Extragalactic Distance Database\footnote{http://edd.ifa.hawaii.edu} (EDD)  \citep{2009AJ....138..323T} website in the 'All Digital HI' catalog.  

A high priority of the project is to increase the size of the sample by both going to larger redshifts and closer to the zone of avoidance while also homogenizing the distribution of distance measurements across the sky. 
This paper gathers original datasets retrieved directly at the archives of Green Bank, Parkes and Arecibo telescopes 
to complement observational programs dedicated to the Cosmic Flows projects. 
Extracted HI profiles are all consistently processed by our linewidth pipeline, which is now also used by the ASKAP-WALLABY collaboration and with the Effelsberg SDSS-Northern Sky Survey. 

The first section of this paper presents the different samples of selected galaxies extracted from the archives. Subsequent sections discuss the data reductions and compilation of linewidths.

\section{Data retrieval}

\subsection{Parkes 64m}

A search for archival material since 2000 obtained with the Parkes-64m telescope led us to review observational programs listed in Table~\ref{PKS}.

\begin{table}
\caption{HI surveys conducted between 2000 and 2010 with Parkes 64m telescope. Data was retrieved from archival disks and explored.
When signal to noise was sufficient, we reduced the data. Column 1 gives the internal Parkes program number. Column 2 gives 
the survey title, investigator's name and date of observation.}
\begin{tabular}{l}
\hline
P248 HIPASS/ZOA Staveley-Smith/Webster (2000)\\
P290 Redshift  behind Milky Way - Saunders-Staveley-Smith (2000)\\
P307 Southern Galactic Plane Survey: Full Survey (Dickey)\\
P312 Narrow-band HI MB Survey of the Magellanic System (Haynes)\\
P335 High Velocity Resolution HI study of Nearby Galaxies (de Blok)\\
P347 HI around dwarf spheroidal galaxies (Carignan)\\
P352 Formation \& Evolution  in Groups - role of HI (Forbes)\\
P357 Northern extension to ZOA (Staveley-Smith)\\
P364 Seibel-Danziger TF of spirals containing SNIa (2005)\\
P370 The baryonic Tully-Fisher  HIPASS sample (Gurovich)\\
P387  Completeness \& reliability of HIPASS (Zwaan)\\
P389  Narrow band observations of HIPASS edge-on spirals (Meyer)\\
P432  HI content of early type galaxies (Garduno)\\
P467  GASS : Galactic all sky survey McClure-griffiths\\
P475  HI content of early-type dwarfs in nearby groups (Bouchard)\\
P561  Mapping Mass Local Universe (Hong, Masters) (2006-2007)\\
\hline
\end{tabular}
\label{PKS}
\end{table}

Programs P364 (PI: Seibel \& Danziger; 69 spectra retrieved) 
and P561 (\cite{2013MNRAS.432.1178H})
(139 spectra retrieved)
produced a total of 142 measurements of linewidths that are now reduced as in  \cite{2011MNRAS.414.2005C} and included in our catalog.
The new material from Parkes can be identified in table~\ref{new} by "archi12" (archive search 2012) in the Source of Information column.

\subsection{GBT 110m}

At GBT four observational programs were studied: GBT06A027,  GBT06B021,  GBT06C049, and  GBT08B003 (see  \cite{2014MNRAS.443.1044M}).
These programs provide observations of 530 2MASS galaxy targets, from which 486 galaxies were cross-identified with a galaxy with a PGC number
\cite{1991rc3..book.....D}, and
 417 had sufficient signal to give an acceptable linewidth measurement. For 44 targets the signal to noise is insufficient for a galaxy distance measurement.
 The new material from the GBT archive can be identified in table~\ref{new} by "archi12" (archive search 2012) in the Source of Information column.

\subsection{Arecibo 300m}

We are following-up on each data release from the extended extragalactic HI blind survey at Arecibo conducted  by Martha Haynes and collaborators.
From 15,855 AGC numbers in \citet{2011AJ....142..170H},  11,941 were cross-identified with a galaxy with a PGC number.
Of these, 1,296 spectra were already analyzed by us following previous ALFALFA releases (1-2-3), leaving us with 9,274 new spectra to
inspect, which we carried out ordered by decreasing signal to noise ratio. The lower 5,000 spectra were left without analysis since the signal to noise was too low
for proper linewidth measurements. This study results in $\sim 3,500$ HI linewidth measurements of which  3,440 are associated with new galaxies to our catalog of HI measurements.
The new material from the Arecibo can be identified in table~\ref{new} by "hgm2011" (Haynes, Giovanelli, Martin 2011) in the Source of Information column.\\

\section{Data reduction} 

The new HI profiles were measured in the same consistent way using "Wmean50". An algorithm written by us in IDL, as described in \cite{2009AJ....138.1938C,2011MNRAS.414.2005C}. 
A $W_{m50}$ parameter corresponds to the HI profile width at $50\%$ of the mean flux within the velocity range encompassing $90\%$ of the total HI flux. This window captures the rotation motions and measures only signal above the noise. In addition, $W_{m50}$ was demonstrated to
provide robust linewidth measurements insensitive to profile shape such as  single peak profiles or asymmetric double peak profiles.\\

Appropriate adjustments of $W_{m50}$ to remove a slight relativistic broadening and a broadening because of finite spectral resolution are applied
before publication of the results: 
 \begin{equation}
W_{m50}^{c} = \frac{W_{m50}}{1+z} - 2 \Delta \nu \lambda
\label{broadening}
\end{equation}
with galaxy heliocentric velocity $cz$, spectral resolution after smoothing $\Delta \nu$ and $\lambda$ determined empirically. Broadening is statistically described with $\lambda = 0.25$.\\
 Linewidths are adjusted to twice maximum rotation velocities $V_{max}$ by a model based upon samples of global profiles and detailed rotation curves 
 (\cite{1985ApJS...58...67T}, \cite{verheijen}):
\begin{equation}
W_{R,\ell}^{2} = W_{\ell}^{2} + W_{t,\ell}^{2} (1 - 2 e^{-(W_{\ell}/W_{c,\ell})^{2}}) - 2 W_{\ell} W_{t,\ell} (1- e^{-(W_{\ell}/W_{c,\ell})^{2}})
\label{adjustment}
\end{equation}
\noindent
with the subscript $\ell$ standing for the specifics of the linewidth estimate, with corresponding turbulent broadening $W_{t,\ell}$, and a transition from boxcar to Gaussian intrinsic profiles characterized by $W_{c,\ell}$. With the current linewidth convention, $W_{c,m50} = 100$~km s$^{-1}$ and $W_{t,m50} = 9$~km s$^{-1}$ give the best fit. The result, $W_{R,m50} \equiv W_{mx}$, is corrected for inclination with a division by sin~$i$ where $i$ is the inclination from face-on. It is defined by the expression ${\rm cos}~i = [\frac{(b/a)^{2}-q_0^2}{1-q_0^2}]^{\frac{1}{2}}$ where $b/a$ is the ratio of minor to major axis and $q_0 = 0.20$ is taken as the reference for a galaxy viewed edge-on \cite{1958MeLu2.136....1H}.\\

Arguments can be made for a more complex dependence of $q_0$. Fortunately, the choice of $q_0$ has a negligible effect on distance measurements if one is consistent between measurements. For example, $q_0$ = 0.13 yields an inclination of $81^{\circ}$ instead of $90^{\circ}$ for $b/a = 0.20$ resulting in a $1/{\rm sin}~i$ difference on the corrected linewidth of only 1.2\%. As one progresses toward larger $b/a$, the difference in assigned inclination is reduced but the $1/{\rm sin}~i$ correction is growing. The product $(b/a)/{\rm sin}~i$ is a roughly constant shift of 1.2\% in the corrected linewidth at all inclinations $i > 45^{\circ}$. Handling every galaxy profile in the same manner will remove any significant effect on the measured distances \citep{2000ApJ...533..744T}.\\

\begin{figure}
\begin{tabular}{l}
%\vspace{-1.4cm}
\includegraphics[width=0.5\textwidth]{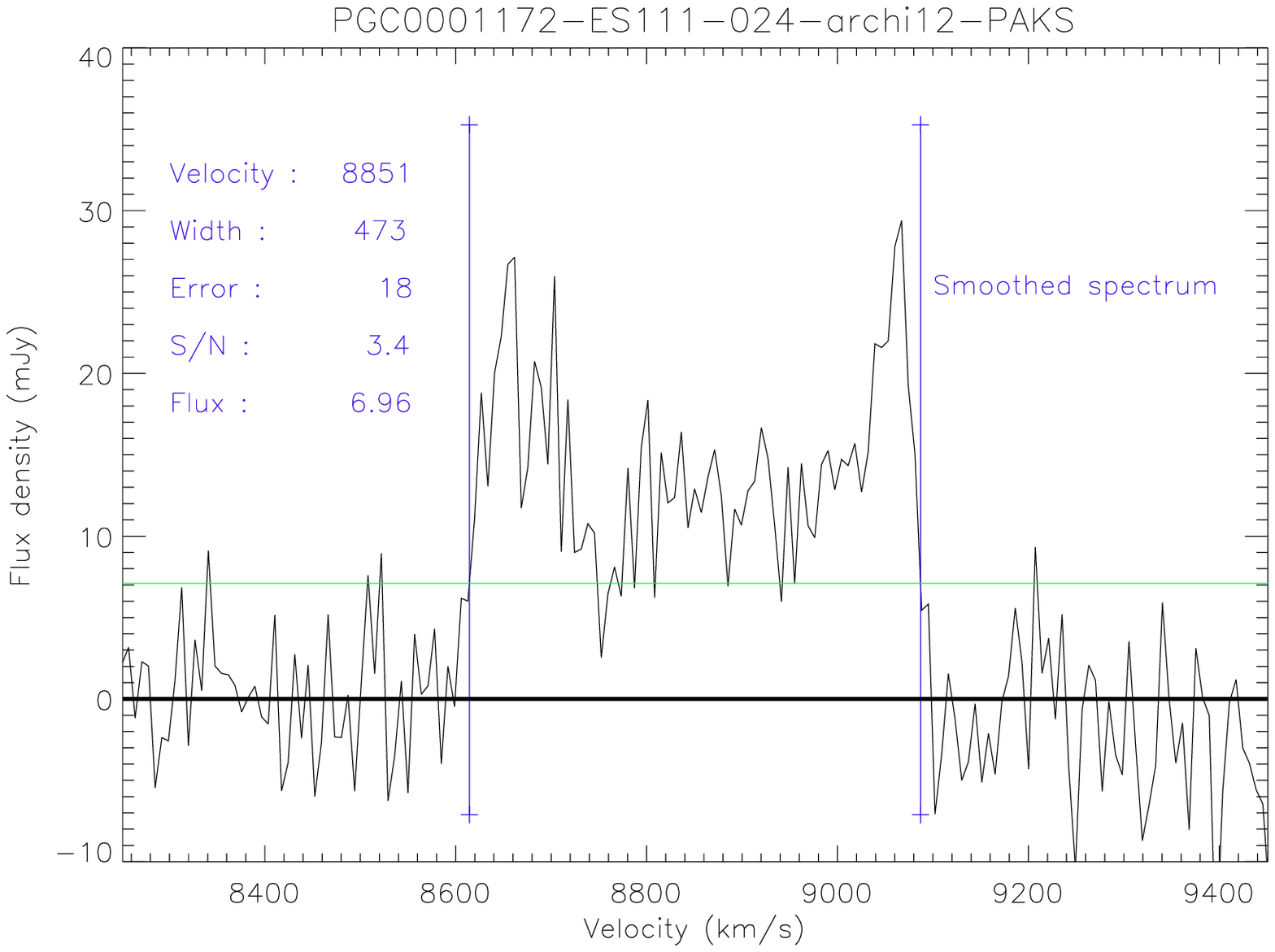}\\
%\vspace{-1.4cm}
\includegraphics[width=0.5\textwidth]{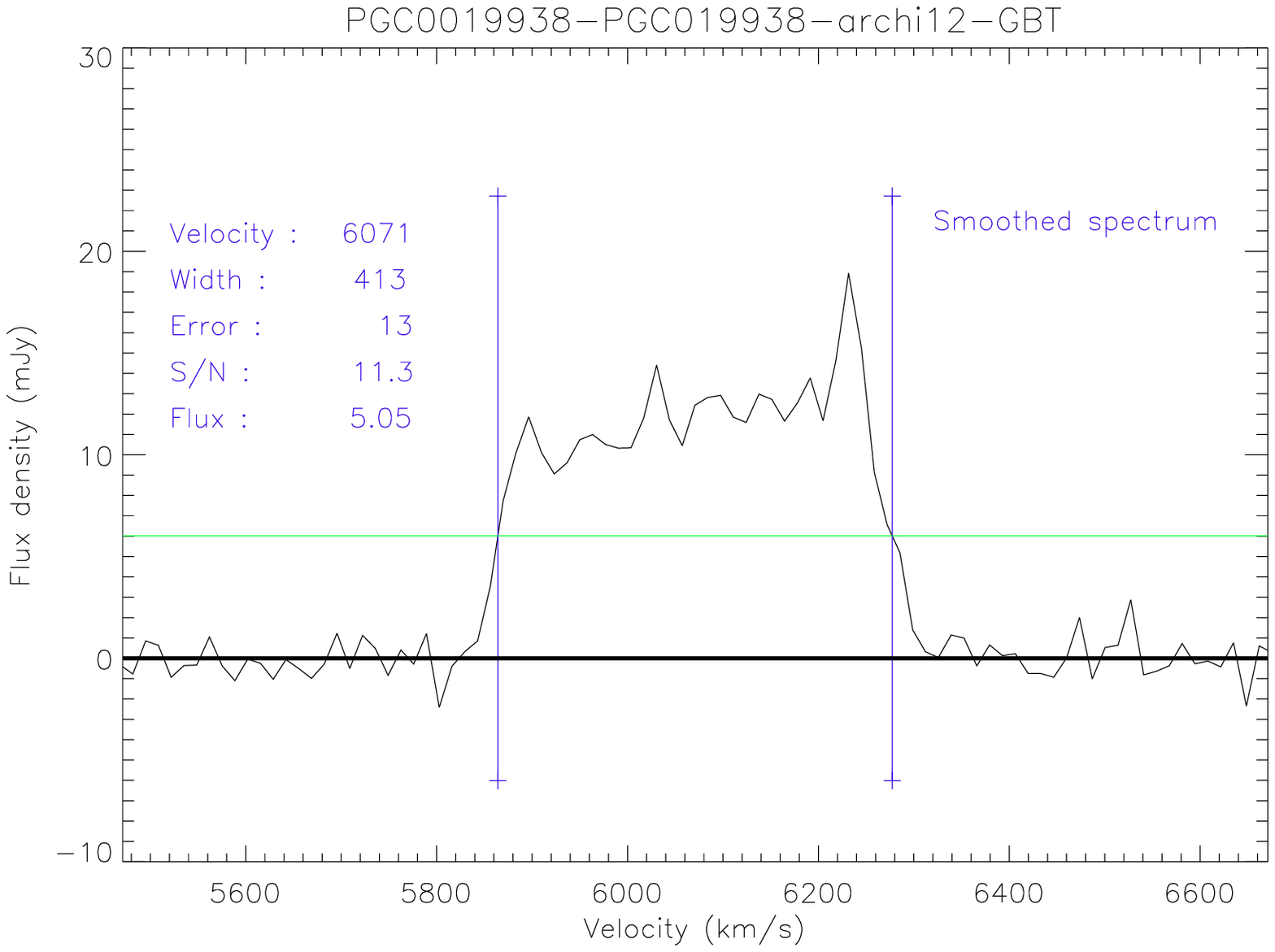}\\
%\vspace{-1.2cm}
\includegraphics[width=0.5\textwidth]{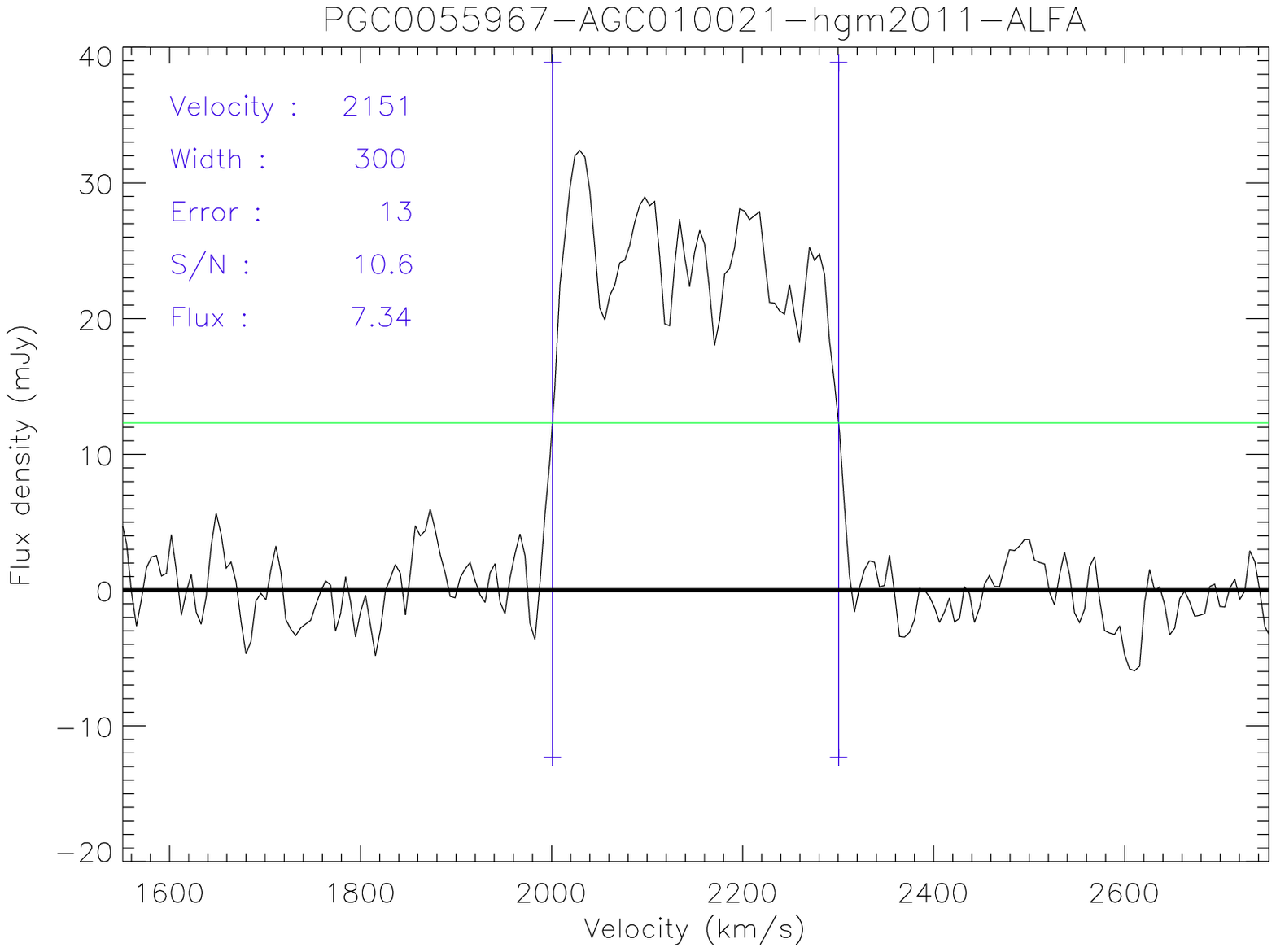}
\end{tabular}
\caption{Examples of three new HI profiles from  Parkes, GBT and Arecibo telescopes (top to bottom). All the 20,343 HI spectra analyzed
for the 17,738 galaxies can be viewed online with the request including the PGC number of the galaxy at the end of the line as one of the galaxy in this example:  for PGC0001172,
type :  "http://edd.ifa.hawaii.edu/get\_profiles.php?pgc=1172" }
\label{ARXIV}
\end{figure}

Width error estimates are based on the level of the signal, S, at 50\% of mean flux over the noise, N, measured beyond the extremities of the signal. Only profiles with error estimates smaller than 20~km s$^{-1}$ are retained for galaxy distance measurements. The corresponding flux per channel is $S/N \geq 2$. 
Examples of new profiles from Parkes, GBT and Arecibo can be seen in Fig.~\ref{ARXIV}. All these
4,117 new HI material measurements are given in an online table with this article, with examples given in Table~\ref{new}.	 
In Table~\ref{new} the telescope acronym PAKS, GBT and ALFA stands respectively for Parkes, Green Bank and Arecibo telescopes.

\section{Discussion and Conclusion}

The Cosmic Flows project started in 2006 by compiling available digital HI spectra for Tully-Fisher measurements of galaxy distances.
In \cite{2011MNRAS.414.2005C}, the HI online catalog, built using "Wmean50" code, contained 16,004 galaxies from which 11,074 had a linewidth of acceptable quality.
Some of these profiles were re-reduced, for example, when a new HI profile coming from a  different telescope gave a discordant value
or, by contrast, was confirming a dubious old measurement. Also some profiles were scrutinized by eye at the time
of matching those radio measurements to $I$-Band, Spitzer or WISE photometry for the purpose of deriving a galaxy distance
(\cite{2011MNRAS.415.1935C}, \cite{2014MNRAS.444..527S}, \cite{2014ApJ...792..129N}).

We thus provide with this article a fully updated catalog of 17,738 galaxies from which 
14,802 have a high enough signal/noise for a distance measurement, an increase of a third over our previous data-release. Table~\ref{good} provides the first few lines of this updated catalog.

\begin{table*}
\caption{The full table is available as online supplementary material. It contains  4,117 galaxies with new additional line width measurements.
 Description of the columns: (1) PGC: Principal Galaxy Catalog name (2) Alternative galaxy name (3) Source code for HI spectrum (4) Telescope acronym (5) $V_{hel}$: heliocentric velocity (6) $W_{m50}$: line width at 50\% of mean flux (7) $W^c_{m50}$: line width at 50\% of mean flux corrected for redshift and spectral resolution broadening (8) $W_{mx}$: line width approximating twice the maximum rotation, not corrected for galaxy inclination; blank if error $>$ 20 km/s (9) $eW$: error in linewidth measure in km/s (10) $S/N$: signal to noise (11) HI line flux in Jy~km/s (12) Spectral channel resolution in km/s (13) $F_{m50}$: flux/channel at 50\% of mean in mJy}
 \begin{tabular}{rrrrrrrrrrrrrr}
\hline
 PGC   &  Name  & Source &Tel   & $V_{hel}$  & $W_{m50}$ & $W^c_{m50}$ & $W_{mx}$ & $eW$ & $S/N$   & Flux    & Res  & $F_{m50}$ \\
             &                 &           &         & km/s  &km/s    & km/s      &km/s  & km/s &          &Jy.km/s& km/s &     mJy \\
 \hline

       1172  &ESO111-024   &  archi12  &  PAKS  &  8851  &   473  &   456   &  447  &  18 &   3.4&    6.96&   1.7     &  7.1 \\
     19938  &PGC019938     &archi12   & GBT    &6071    & 413    & 398     &389   & 13  & 11.3   & 5.05   &3.3      & 6.0\\
      55967  &AGC010021    & hgm2011 &  ALFA &   2151 &    300  &   295  &   286 &   13 &  10.6 &   7.34&   5.2   &   12.3 \\
.........\\
\hline
\end{tabular}
\label{new}
\end{table*}

\begin{table*}
\caption{The full table is available as online supplementary material. It contains the averaged measurements for 14,802 galaxies with high quality spectrum.
Description of the columns: (1) PGC: Principal Galaxy Catalog name (2) Alternative galaxy name (3) $V_{hel}$: averaged heliocentric velocity (4) $W_{mx}$: weighted averaged line width approximating twice the maximum rotation, not corrected for galaxy inclination (5) $eW$: error in linewidth measure after averaging in km/s (6) $N$: number of averaged independent measurements (7) $S/N$: signal to noise (10) HI line flux in Jy~km/s }
\begin{tabular}{rrrrrrrrrrrrr}
\hline

PGC     & Other name & $V_{hel}$ &  $W_{mx}$ & $eW$ & $N$ &  $S/N$  &  Flux \\
             &                        &     km/s         &  km/s                 & km/s                       &                  &           &Jy~km/s \\
              \hline
       4  &AGC331060   &  4458    & 154    &16  & 1    &  8.5  &  1.85 \\
       6  &AGC331061    &6002     &217    &20   &1      &2.0   & 0.82 \\
      12  &PG0000012    &6548     &400    &19   &1      &2.4   & 3.40\\
      16  & PG0000016    &5668     &296    &20   &1      &2.2  &  1.04\\
      38   &UGC12893    &1108      &78    &19   &1      &3.8    &2.41\\
      40  &PG0000040    &7282     &289    &20   &1     & 5.0   & 5.20\\
.........\\
\hline
\end{tabular}
\label{good}
\end{table*}

The importance of building large coherent datasets is also manifested in improvements in our
comprehension of galaxy physics, for example with the new calibration of the baryonic Tully-Fisher relation \citep{zbtf}.
In the near future,  teams will be able to measure consistently with sufficient signal, linewidths of tens of thousands of galaxies by
combining output from the multi-dish telescopes: southern ASKAP-Wallaby and northern HI Westerbork surveys WNSHS (\cite{2012MNRAS.426.3385D}) .

These Tully-Fisher measurements are complemented by Fundamental Plane distances for elliptical galaxies
(\cite{2014MNRAS.445.2677S}).
In the future, the TAIPAN survey (\cite{2014SPIE.9147E..10K}),
starting 2015 at the UK Schmidt telescope in Australia, will
 measure 500,000 redshifts to z $\sim$ 0.2 with r $\sim$17, K $\sim$14 and about 10,000 galaxy distances with typical
 errors of the order of 20$\%$.

\section*{Acknowledgements}   

We thank Bryson Lee for his work during a summer internship 2013 at the Institute for Astronomy.
HC  acknowledges support from the Lyon Institute of Origins under grant ANR-10-LABX-66 and from CNRS under PICS-06233.
RBT acknowledges support from the US National Science Foundation award AST09-08846 and NASA award NNX12AE70G. 
We acknowledge the usage of the HyperLeda database (http://leda.univ-lyon1.fr).
This research has made use of the NASA/IPAC Extragalactic Database (NED) which is operated by the
Jet Propulsion Laboratory, California Institute of Technology, under contract with the National Aeronautics and Space Administration.

\label{lastpage}
\clearpage
\bibliography{biblicomplete}
\clearpage
\end{document}